\documentclass[final]{elsart}
\newcommand{\avrg}[1] {\langle  #1  \rangle}

\usepackage{epsfig}
\usepackage{slashbox}

\renewcommand{\d}{{\rm d}}

\usepackage{amssymb}
\usepackage{amsmath}

\begin{document}

\begin{frontmatter}



\title{Simultaneous Determination of Signal and Background Asymmetries}


\author{J\"org Pretz \corauthref{cor1}}
 \corauth[cor1]{corresponding author}
\ead{jorg.pretz@cern.ch}
\address{Physikalisches Institut, Universit\"at Bonn, 53115 Bonn, Germany}
\author{Jean-Marc Le Goff }
\address{CEA Saclay IRFU/SPhN, 91191 Gif-sur-Yvette, France}

\begin{abstract}
This article discusses the determination of asymmetries.
  We consider a sample of events consisting of a peak of signal events on 
  top of some background events. Both signal and background
  have an unknown asymmetry, e.g. a spin or forward-backward asymmetry.
A method is proposed which determines signal and background asymmetries
simultaneously using event weighting.
For vanishing asymmetries the statistical error of the asymmetries reaches the 
minimal variance bound (MVB) given by the
Cram\'er-Rao inequality and it is very close to it for large asymmetries.
The method thus provides a significant gain in statistics compared
to the classical method of side band subtraction of background asymmetries. 
It has the advantage with respect to the unbinned maximum likelihood
approach,
reaching the MVB as well,
that it does not require loops over the event sample
in the minimization procedure.
\end{abstract}

\begin{keyword}
event weighting \sep minimal variance bound \sep Cram\'er-Rao inequality \sep asymmetry  extraction
\sep optimal observables \sep side band subtraction 

\PACS 02.70.Rr \sep 13.88.+e 
\end{keyword}
\end{frontmatter}


\section{Introduction}

\begin{figure}
\includegraphics[width=\textwidth]{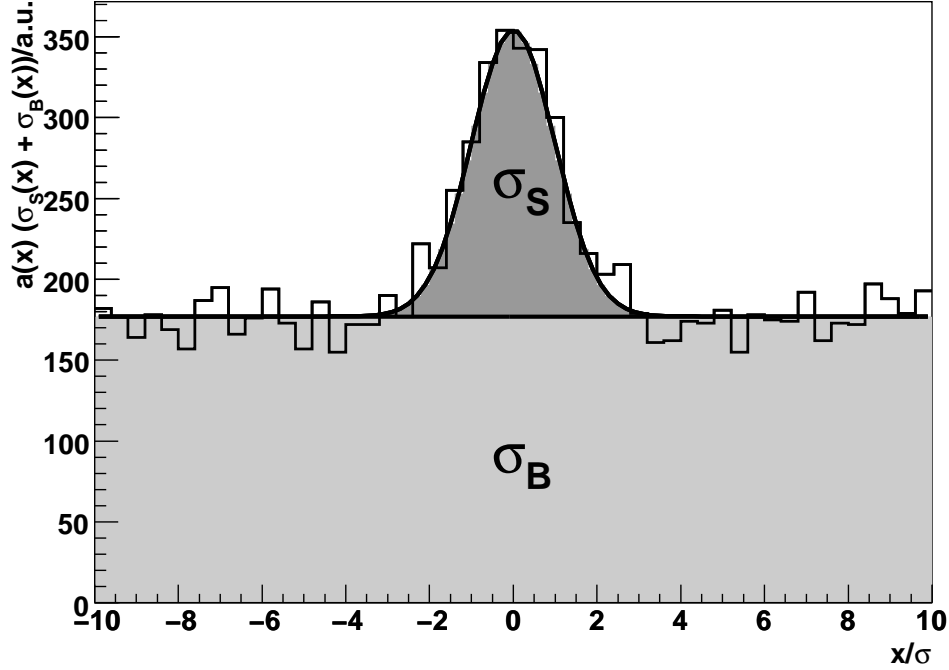}
\caption{Example of signal events originating from a Gaussian distribution
  centered at $x=0$ and width $\sigma=1$ sitting on a constant background.
  \label{massplot}}
\end{figure}

Asymmetries of cross sections,
e.g.~spin-asymmetries and  forward-backward asymmetries, are often
interesting physics quantities.
For concreteness let us consider a situation as shown in Fig.~\ref{massplot},
where the asymmetry of the signal events, in the central Gaussian peak with a width of
$\sigma$, should be 
determined from data taken in two different spin configurations.
The number density of events as a function of some kinematic variable, $x$, 
(typically a reconstructed mass) is given by
\begin{equation*}\label{N1}
  n^\pm(x) = a(x) (\sigma_S(x) + \sigma_B(x)) 
\left( 1 \pm A_S \frac{\sigma_S(x)}{\sigma_S(x) + \sigma_B(x) }
         \pm A_B \frac{\sigma_B(x)}{\sigma_S(x) + \sigma_B(x) }
 \right) 
\end{equation*}  
with
$\sigma_{S,B} = \frac{1}{2} (\sigma_{S,B}^+ + \sigma_{S,B}^-)$.
Here $\sigma_S^{\pm}$ ($\sigma_B^{\pm}$) denotes the cross section of the
signal (background) events in the two different spin configurations $+$ and $-$. %
The factor $a$ is a luminosity and acceptance factor, 
assumed to be the same for the two spin configurations.
The goal is to determine from spectra as shown in Fig.~\ref{massplot},
and taken in two spin configurations, the two unknown asymmetries 
$A_{S} = (\sigma_{S}^+ - \sigma_{S}^-)/(\sigma_{S}^+ + \sigma_{S}^-)$  
and $ A_{B} = (\sigma_{B}^+ - \sigma_{B}^-)/(\sigma_{B}^+ + \sigma_{B}^-)$,
assumed to be independent of $x$.
It is of course not known event-by-event whether
a particular  event is signal or background; one only knows
the fraction of signal events as a function of $x$,
from a fit to the event spectrum as in Fig.~\ref{massplot}.

Section~\ref{simple} presents
the simplest method, based on counting rate asymmetries. 
Section~\ref{lh} describes the unbinned likelihood method 
which is known to yield the smallest possible variance of all unbiased
estimators in the limit of an infinite number of events, 
thus reaching the minimal variance bound  (MVB) given by the Cram\'er-Rao inequality.
Section~\ref{weighting} presents a new asymmetry estimator, based on weighted 
events. This estimator is also unbiased in the large $N$ limit, i.e.~it is
consistent, and it is very close to reach the minimal variance bound. The advantage is that it can also be used 
in cases where the unbinned likelihood method is cumbersome because
of large number of events.
Event weighting to extract the number of signal and background
events was discussed in Ref.~\cite{barlow1} but extraction of
asymmetries is not discussed in this reference.
The different methods are compared in section~\ref{dis}.

\section{Estimator based on counting rate asymmetries }\label{simple}

A method often found in the literature \cite{Barate:1998yi,Adare:2007dg} is to determine the 
asymmetry in a $k$-standard-deviation region around the peak, a region which 
includes both signal and background; then to measure   
the background asymmetry in some side bands around the signal peak 
($-k_{max}\sigma<x<-k_{min}\sigma$ and $k_{min}\sigma<x<k_{max}\sigma$) and to
use the result to correct the asymmetry measured in the peak region. 
For sake of simplicity we will set $\sigma=1$, 
so that everywhere below we can write $k$ instead of $k\sigma$.

The expectation value of the counting rate asymmetry, $A^{cnt}$,  in the range $-k< x <k$  is related
to $A_S$ and $A_B$ in the following way:
\begin{equation}
\label{Eq:<A_cnt>}
  \avrg{A^{cnt}} = \frac{ \avrg{N^+} - \avrg{N^-}}{\avrg{N^+} + \avrg{N^-}} 
= A_S \frac{\int_{-k}^{k} a \sigma_S \d x}{\int_{-k}^{k} a (\sigma_S + \sigma_B) \d x} +
                                         A_B \frac{\int_{-k}^{k} a \sigma_B \d x}{\int_{-k}^{k} a (\sigma_S + \sigma_B) \d x} \, , 
\end{equation}
where we used $\avrg{N^+} = \int n^+(x) \d x $ and $\avrg{N^-}= \int n^-(x) \d x$. 
An estimator for $A_S$ is given by:
\begin{equation}
\label{Eq:tilde-A_S}
\tilde A_S = \frac{\int_{-k}^{k} a (\sigma_S+\sigma_B) \d x}{\int_{-k}^{k} a \sigma_S \, \d x} 
\left( {A^{cnt}} - \frac{\int_{-k}^{k} a \sigma_B \d x}{\int_{-k}^{k} a (\sigma_S +\sigma_B) \d x} A_B \right) \, .
\end{equation}
Note that, strictly speaking, the first equality in Eq.~(\ref{Eq:<A_cnt>}) is valid only in the large $N$ limit. In this limit Eqs.~(\ref{Eq:<A_cnt>}) and
(\ref{Eq:tilde-A_S}) indicate that $\avrg{\tilde A_S}=A_S$, i.e.~$\tilde A_S$ is a consistent estimator.

The corresponding figure of merit, FOM=$1/\sigma^2_{\tilde A_S}$, 
reads
 \begin{equation}\label{fom1}
    \mbox{FOM} =   \left(\frac{\int_{-k}^{k} a \sigma_S \d x}
         {\int_{-k}^{k} a (\sigma_S + \sigma_B) \d x} \right)^2  
       \left( \sigma^2_{A^{cnt}} + \left( \frac{\left(\int_{-k}^{k} a \sigma_B \d x \right)}{\int_{-k}^{k} a (\sigma_S + \sigma_B) \d x} \right)^2 \, 
        \sigma^2_{A_B} \right)^{-1} .
\end{equation}
Here and in the following we assume small asymmetries,
such that for the error calculation the approximation 
$\avrg{N^+} \approx \avrg{N^-}$ is valid.
In this case one finds
$  1/\sigma^2_{A^{cnt}} = \int_{-k}^{k} \left[n^{+}(x) + n^{-}(x)\right]  \d x$  and
  $1/\sigma^2_{A_B}  = \int_{-k_{max}}^{-k_{min}} [n^{+}(x) + n^{-}(x)]  \d x  
                    +  \int_{k_{min}}^{k_{max}} [n^{+}(x) + n^{-}(x)]  \d x $.
Introducing these values of $\sigma^2_{A^{cnt}}$ and $\sigma^2_{A_B}$ in  Eqs.~\ref{fom1} shows that the FOM depends
on the choice of both the signal region ($k$) and the background region ($k_{min}$ and $k_{max}$).
The solid line in  Fig.~\ref{fom} shows the FOM as a function of $k_{max}$, for $k_{min}=3$ which is a reasonable value to make sure that the side bands include a negligible amount of signal.
The signal region, i.e.~the value for $k$, is chosen in order to maximize the
FOM for the given $k_{max}$.
The FOM depends also on the signal to background ratio,
here chosen to be 1:1 at $x=0$, as in Fig.~\ref{massplot}.

\section{Maximum Likelihood asymmetry estimators}\label{lh}

In the large $N$ limit, the unbinned maximum likelihood method is known to provide an unbiased estimator for the
parameters $A_S$ and $A_B$, which reaches the minimal variance bound.
Since the numbers of events $N^+$ and $N^-$ are not fixed, an extended maximum likelihood method
has to be used~\cite{barlow}. 
With the definitions 
$   S_i = \sigma_S(x_i)/(\sigma_S(x_i) + \sigma_B(x_i))$,
   $B_i = \sigma_B(x_i)/(\sigma_S(x_i) + \sigma_B(x_i))$ and
  $\alpha_i = a(x_i) \, (\sigma_S(x_i) + \sigma_B(x_i))$ 
the log likelihood function reads:
\begin{eqnarray}
  l = \ln \mathcal{L} &=& \sum_1 \ln \left(\alpha_i (1 + S_i A_S + B_i A_B) \right)- \langle N^+\rangle (A_S,A_B)  \nonumber \\
                  &&+ \sum_2 \ln \left(\alpha_i (1 - S_i A_S - B_i A_B) \right)- \langle N^-\rangle (A_S,A_B)  \, ,\nonumber
\end{eqnarray} 
where $\Sigma_1$ ($\Sigma_2$) runs over all events in the $+$ ($-$)
configuration and in the range $-k_{max}<x<k_{max}$,
while $\langle N^\pm\rangle (A_S,A_B) = \int n^{\pm}(x) \d x$.
The first derivative is
\begin{eqnarray}
 \frac{\partial l}{ \partial A_S } &=& \sum_1 \frac{S_i}{1+S_i A_S + B_i A_B} - \sum_2 \frac{S_i}{1 - S_i A_S - B_i A_B} \, ,
\end{eqnarray} 
with a similar expression for $A_B$. Note that the terms with  $\langle N^+
\rangle$ and $\langle N^- \rangle$ cancel each other because 
the same $a$ is assumed for the two configurations. 
The set of equations $\partial l/\partial A_{S,B} = 0$ can be solved for $A_S$ and $A_B$.

For small asymmetries a first order expansion in $A_S$ and $A_B$ gives the set
of equations
\begin{eqnarray}
\label{likelihood-set-equations}
\left(\sum_1 S_i^2 + \sum_2 S_i^2\right) A_S + 
\left(\sum_1 S_i B_i + \sum_2 S_i B_i\right) A_B & 
= &  \sum_1 S_i - \sum_2 S_i	\, , \nonumber \\
\left(\sum_1 S_i B_i + \sum_2 S_i B_i\right) A_S 
+ \left(\sum_1 B_i^2 + \sum_2 B_i^2\right) A_B 
& = &  \sum_1 B_i - \sum_2 B_i
\end{eqnarray}
and the covariance matrix of the two parameters $A_S$ and $A_B$ reads :
\begin{equation}
  \mbox{cov}^{-1}(A_S,A_B) = 
-\left(
\begin{array}{cc} 
\frac{\partial^2 l}{\partial A_S^2} & \frac{\partial^2 l}{\partial A_S \partial A_B} \\
 \frac{\partial^2 l}{\partial A_S \partial A_B} & \frac{\partial^2 l}{\partial A_B^2} \\
\end{array}
\right)
=
\left(
\begin{array}{cc} 
\sum S_i^2 & \sum S_i B_i \\
 \sum S_i B_i & \sum B_i^2 \\
\end{array}
\right) \, ,
\end{equation}
For the FOM of $A_S$ one finds
\begin{eqnarray}
  \mbox{FOM} = (1-\rho^2) \sum S_i^2  \quad \mbox{with} \quad \rho = \frac{\sum S_i B_i}{\sqrt{\sum S_i^2 \sum B_i^2 }} \, .
 \end{eqnarray}
Note that, if not otherwise stated, all sums run over both event samples, 1 and 2.

\begin{figure}
\includegraphics[width=\textwidth]{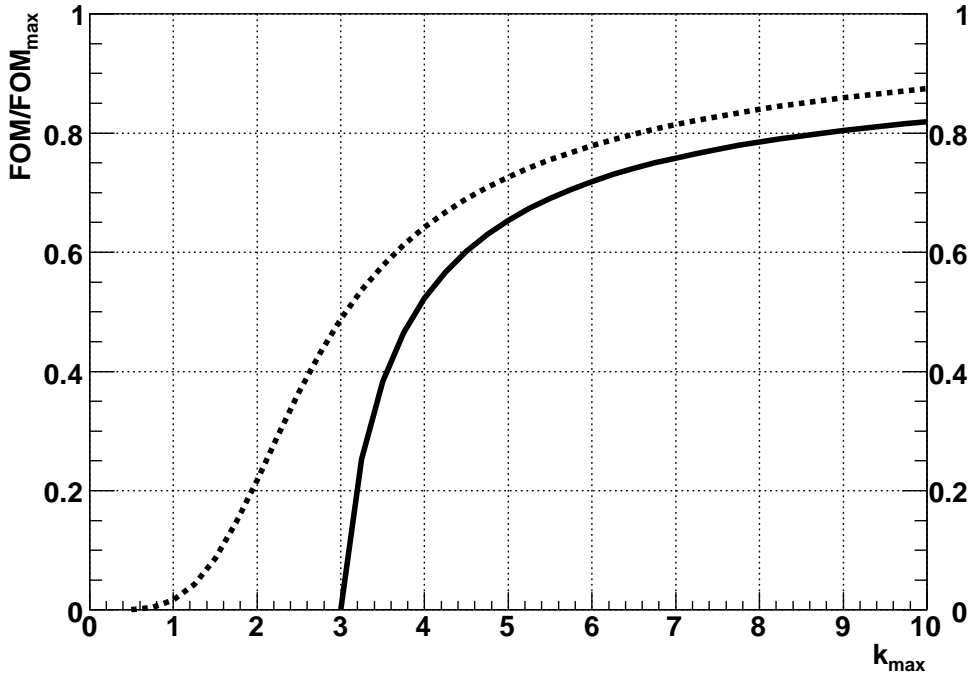}
\caption{ 
FOM of $A_S$ as a function of the maximum range of data available defined by $k_{max}$,
for the classical method of side band subtraction (solid line) and
for the likelihood or weighting method (dotted line).
In the side band subtraction method $k_{min}=3$ and for each value of
$k_{max}$ the value of $k$, defining the signal region, is chosen in order to maximize the FOM.
The figures of merit are normalized to the maximum FOM reachable in the likelihood or weighting method
in the limit $k_{max}\rightarrow \infty$. In this case $\mbox{FOM} = \sum_{1,2} S_i^2$.
 \label{fom}}
\end{figure}

The dotted line in Fig.~\ref{fom} shows this FOM as a function of $k_{max}$,
i.e. for events in the region $-k_{max} < x < k_{max}$. For a given range of
data available, defined by $k_{max}$,
it is always larger than the FOM obtained with the side band subtraction method
shown by the solid line. The latter method does not reach the minimal variance bound.

\section{Extracting the asymmetries using event weighting}\label{weighting}

In this section a method to extract $A_S$ 
(and simultaneously $A_B$) using event weighting is developed.
It is clear that the estimator based on the counting rate asymmetry is not statistically optimal since it gives the same weight to all events.
Better estimators can be obtained by weighting each event by the
signal strength, $S_i$, and by the background strength, $B_i$. 
These weight factors coincide with the optimal
weights found in~\cite{barlow1} to extract the number of signal events.
They are used to build the following asymmetries:
\begin{equation}
  a_S = \frac{\sum_1 S_i - \sum_2 S_i}{\sum_1 S_i + \sum_2 S_i} \; , \quad \; \;
  a_B = \frac{\sum_1 B_i - \sum_2 B_i}{\sum_1 B_i + \sum_2 B_i}  \, .
\end{equation}  
In the large $N$ limit, the expectation values of $a_S$ and $a_B$ are
\begin{eqnarray}
    \avrg{a_S} &=& A_S \frac{\int \alpha S^2 \d x}{\int \alpha S \d x} 
                + A_B \frac{\int \alpha B S \d x}{\int \alpha S \d x}   \, ,\label{as}\\
    \avrg{a_B} &=& A_S \frac{\int \alpha B S \d x}{\int \alpha B \d x} 
                + A_B \frac{\int \alpha B^2 \d x}{\int \alpha B \d x}   \, , \label{ab}
\end{eqnarray}
where $\alpha=a(x) \, (\sigma_S(x) + \sigma_B(x))$, as in section~\ref{lh}.
 The ratios of integrals can easily be obtained from the event sample, e.g.
$  \int \alpha S^2 \d x \, / \, \int \alpha S \d x  \approx \sum_{1,2} S_i^2 \, / \sum_{1,2} S_i$, which results exactly in the set of equations (\ref{likelihood-set-equations}) found for the 
likelihood method in the small asymmetry limit.
So the FOM is still 
$1/\sigma_{A_S}^2 =  (1-\rho^2) \sum S_i^2 $. 
This result can of course also be obtained directly, by simple error propagation
using the expressions found for $A_S$ and $A_B$ from Eqs.~(\ref{as}) and (\ref{ab}). 
Appendix~\ref{cov}
shows that  
the factor $\rho$ is actually the correlation coefficient 
between $\sum S_i$ and $\sum B_i$.

This shows that the weighting method and the unbinned 
likelihood method are identical for small asymmetries.
The advantage of the weighting method is that the estimators
derived from Eqs.~(\ref{as}) and (\ref{ab}) can also be used for arbitrary asymmetries,
whereas the likelihood method requires in this case 
a numerical maximization of $\ln \mathcal{L}$ with loops
over all events.
For sake of simplicity, the error calculation was only 
presented for small asymmetries. Extending it to arbitrary
asymmetries is straightforward but lengthy;
it shows that the FOM of the weighting method is only slightly 
smaller than the FOM of the unbinned LH method.
For example for a signal to background ratio as given in
Fig.~\ref{massplot} and asymmetries smaller than 50\% the decrease
in the FOM is less than 1\%.

The weighting method can also be extended to more complicated cases
where for example the acceptance factors $a$ are not the same in the two spin
configurations or even
when the asymmetries have to be determined from four counting rates in order
to cancel differences of acceptances and flux factors for the two spin configurations, as in Ref.~\cite{compass_oc}. 

\section{Discussion of the results \& summary}\label{dis}

A comparison of the two curves in Fig.~\ref{fom} shows that the FOM 
of the likelihood or event weighting method
is always larger than the corresponding FOM for the classical method.
For a signal-to-background ratio of 1:1 at $x=0$, as in Fig.~\ref{massplot}, the gain is 23\% for $k_{max}=4$ and 7\% for $k_{max}=10$. 
For $k_{max}=10$ the gain is 2\% and 10\% for a signal-to-background ratio
of 10:1 and 1:10, respectively.
Apart from the gain in statistics it should also be noted that the weighting method
avoids the arbitrary choice of the background region
which starts here at $3\sigma$. For Breit-Wigner distributions for example 
this choice is less obvious.

In summary, a new set of two estimators was presented to determine simultaneously signal and background asymmetries. 
These estimators are unbiased in the large $N$ limit,
i.e.~they are consistent. For small asymmetries they are also efficient, i.e.~they reach 
the minimal variance bound, like the statistically optimal unbinned likelihood method. 
This is in contrast to the classical method of side band subtraction.
These estimators can actually be derived from the likelihood method in the case of vanishing asymmetries. 
For large asymmetries their variances are still very close to the minimal variance bound.
The advantage of the method is its applicability in cases where the likelihood method is cumbersome.

\appendix
\section{Derivation of the covariance matrix $\mbox{cov}(a_S,a_B)$ and
  correlation coefficient $\rho$} \label{cov}
Consider two weight factors $S$ and $B$.
The covariance between $\sum_i S_i$ and $\sum_j B_j$ is given by:
\begin{eqnarray*}
   \lefteqn{\mbox{cov}(\sum_i S_i, \sum_j B_j)}\\  
   &=& \avrg{ \sum_i S_i \sum_j B_j} - \avrg{ \sum_i S_i} \avrg{\sum_j B_j}   \\
                                  &=& \avrg{ \sum_{i=j} S_i B_i + \sum_{i\ne j} S_i B_j} 
                                       - \avrg{ \sum_i S_i} \avrg{\sum_j B_j}  \\
                                   &=& \avrg{N}  \avrg{S B} + \avrg{N(N-1)}  \avrg{S}  \avrg{B} - \avrg{N}^2 \avrg{ S} \avrg{B}  \\
                                   &=& \avrg{N}  \avrg{S B} + (\avrg{N^2} - \avrg{N} - \avrg{N}^2)  \avrg{S}  \avrg{B}   \, .
\end{eqnarray*}
If the number of events $N$ is Poisson distributed, i.e.
$\avrg{N^2} - \avrg{N} - \avrg{N}^2 =0$,
one finds cov$(\sum_i S_i, \sum_j B_j) = \avrg{N}  \avrg{S B}  \approx \sum_i S_i B_i$.
The error on the sums of weights is given by
$\sigma^2_S = \mbox{cov}(\sum_i S_i, \sum_j S_j) =  \sum_i S_i^2$ .
Thus the correlation coefficient is
\begin{equation}\label{rho}
  \rho = \frac{\mbox{cov}(\sum_i S_i, \sum_j B_j)}{\sigma_S \sigma_B} 
     = \frac{\sum_i S_i B_i}{\sqrt{ \sum_i S_i^2 \sum_i B_i^2 }} \; . 
\end{equation}

\label{}



\end{document}